\documentclass[aps,twocolumn,secnumarabic,groupedaddress,floatfix,amsfonts]{revtex4}
\usepackage[dvips]{graphicx,color}
\begin{document}
\title{Data set diagonalization in a global fit}
\author{Jon Pumplin}
\affiliation{
Michigan State University, East Lansing MI 48824, USA 
}
%\email{pumplin@pa.msu.edu}
\date{\today}
\preprint{MSUHEP-090416}
\begin{abstract} 
The analysis of data sometimes requires fitting many free parameters in a 
theory to a large number of data points.  Questions naturally arise about 
the compatibility of specific subsets of the data, such as those from a 
particular experiment or those based on a particular technique, with the 
rest of the data.  Questions also arise about which theory parameters are 
determined by specific subsets of the data.  I present a method to answer 
both of these kinds of questions.  The method is illustrated by applications 
to recent work on measuring parton distribution functions.
\end{abstract}
%\pacs{12.38.Qk, 12.38.Bx, 13.60.Hb, 13.85.Qk}
\maketitle

\section{Introduction}
\label{sec:introduction}

There are many situations where data from a variety of different experiments 
must be fitted to a single underlying theory that has many free parameters.  
The particular instance that led to this work is the measurement of parton 
distribution functions (PDFs), which describe momentum distributions of 
quarks and gluons in the 
proton \cite{HessianMethod,LagrangeMethod,cteq66,CT09,MSTW08}.

In these situations, it would be desirable to assess the consistency between 
the full body of data and individual subsets of it, such as data from a 
particular experiment, or data that rely on a particular technique, or data 
in which a particular kind of theoretical or experimental systematic error 
is suspected.  It would also be desirable to characterize which parameters 
in the fit are determined by particular components of the input data.  
This paper presents a ``Data Set Diagonalization'' (DSD) procedure that 
answers both of those desires.

\section{New eigenvector methods}
\label{sec:EigenvectorMethods}

The quality of the fit of a theory to a set of data is measured 
by a quantity\ $\chi^2$, which in simplest form is given by
\begin{equation}
\chi^2 \, = \, \sum_{i=1}^M \left(\frac{D_i -
T_i}{E_i}\right)^2 \, ,
  \label{eq:chisqdef}
\end{equation}
where $D_i$ and $E_i$ represent a data point and its uncertainty, and $T_i$ 
is the theoretical prediction. (Although (\ref{eq:chisqdef}) is standard 
practice, some alternatives might be worth consideration \cite{alternatives}.)

The predictions $T_i$ in Eq.\ (\ref{eq:chisqdef}) depend on a number of 
parameters $a_1,\dots,a_N$.  
The best-fit estimate for those parameters is obtained by adjusting them to 
minimize $\chi^2$.  The uncertainty range is estimated as the neighborhood 
of the minimum in which $\chi^2$ lies within a certain 
``tolerance criterion'' $\Delta \chi^2$ above its minimum value.  If the errors 
in the data are random and Gaussian with standard deviations truly given by 
$E_i$, and the theory is without error, the appropriate $\Delta \chi^2$ 
can be related to confidence intervals by standard statistical methods.
Those premises do not hold in the application of interest here; but the 
tolerance range can be estimated by examining the stability of the fit in 
response to applying different weights to subsets of the 
data \cite{HessianMethod,LagrangeMethod,Collins}.

Sufficiently close to its minimum, $\chi^2$ is an approximately quadratic 
function of the parameters $a_1,\dots,a_N$.  Using the eigenvectors of 
the matrix that defines that quadratic form as basis vectors in the 
$N$-dimensional parameter space, one can define new theory parameters 
$z_1,\dots,z_N$ which are linear combinations of the original ones
\begin{equation}
a_i \, = \, a_i^{(0)} \, + \, \sum_{j=1}^N W_{ij} \, z_j \; ,
\label{eq:lineartrans} 
\end{equation}
and which transform $\chi^2$ into the very simple form 
\begin{equation}
  \chi^2 \, =  \, \chi_{\mathrm{min}}^{\,2} \, + \, 
               \sum_{i=1}^N z_i^{\, 2} \; . 
  \label{eq:diagchi} 
\end{equation}
Formally, the transformation matrix $\mathbf{W}$ can be computed by evaluating 
the Hessian matrix $\partial^{\, 2} \, \chi^2 / \partial a_i \, \partial a_j$ 
at the minimum using finite differences, and computing its eigenvectors.  
The new parameters $z_i$ are then just coefficients that multiply those 
eigenvectors when the original coordinates $a_1,\dots,a_N$ are expressed as 
linear combinations of them.
In the PDF application, this straightforward procedure breaks down 
because the eigenvalues of the Hessian span a huge range of magnitudes,
which makes non-quadratic behavior complicate the finite-difference 
method at very different scales for different directions in parameter space.  
However, this difficulty can be overcome by an iterative 
technique \cite{HessianMethod} that is reviewed in the Appendix.

The linear transformation (\ref{eq:lineartrans}) that leads to 
(\ref{eq:diagchi}) is not unique, since any further orthogonal transform of the 
coordinates $z_i$ will preserve that form.  Such an orthogonal transformation 
can be defined using the eigenvectors of any symmetric matrix.  After this 
second linear transformation of the coordinates, the chosen symmetric matrix 
will be diagonal together with $\chi^2$.  
The second transformation can be combined with the first to yield a 
single overall linear transformation of the form (\ref{eq:lineartrans}).
Thus there is a freedom to diagonalize an additional symmetric matrix while 
maintaining the simple form (\ref{eq:diagchi}) for $\chi^2$.  

That symmetric matrix can be taken from the matrix of second derivatives 
that appears when the variation of any function of the fitting parameters is 
expanded in Taylor series through second order.  \emph{Thus it is possible 
within the quadratic approximation to diagonalize any one chosen function 
of the fitting parameters, while maintaining the diagonal form for $\chi^2$.}  
An explicit recipe for this ``rediagonalization'' procedure is given in the 
Appendix.  

The freedom to diagonalize an additional quantity along with 
$\chi^2$ can be exploited in several ways:
\begin{enumerate}
\item
The traditional approach in which one only diagonalizes the Hessian matrix is 
formally equivalent to also diagonalizing the displacement distance $D$ 
from the minimum point in the space of the original fitting parameters:
\begin{equation}
D^2 = \sum_{i=1}^N \, (a_i -a_i^{(0)})^2 \; .
\label{eq:displacement} 
\end{equation}
In this approach, the final eigenvectors can usefully be ordered 
by their eigenvalues, from ``steep'' directions in which $\chi^2$ rises 
rapidly with $D$, to ``flat'' directions in which $\chi^2$ varies very 
slowly with $D$.  This option has been used in the iterative method that 
was developed for previous CTEQ PDF error analyses \cite{multivariate}.
\item
One can diagonalize the contribution to $\chi^2$ from any chosen subset 
$\mathbf{S}$ of the data.  
This option is the basis of the DSD procedure, which is described in the 
next Section and applied in the rest of the paper.
\item
One can diagonalize some quantity $G$ that is of particular theoretical 
interest, such as the prediction for some unmeasured quantity.  In this 
way, one might find that a small subset of the eigenvectors is responsible 
for most of the range of possibilities for that prediction, which would 
simplify the application of the Hessian method.  An example of this was 
given in a recent PDF study \cite{CT09}.  However, there is no guarantee 
in general that the diagonal form will be dominated by a few directions 
with large coefficients ($\beta_i$ and/or $\gamma_i$ in 
Eq.~(\ref{eq:MainResult}) of the Appendix).  Hence a better scheme to reduce 
the number of important eigenvectors might well be to simply choose the new 
$z_1$ along the gradient direction $\partial G / \partial z_i$, and then to 
choose the new $z_2$ along the orthogonal direction that carries the largest 
residual variation, etc. 
\end{enumerate}

\section{The DSD method}
\label{sec:DSD}

Let us diagonalize the contribution $\chi_{\mathbf{S}}^{\, 2}$
from some chosen subset $\mathbf{S}$ of the data.  
That puts its contribution to the total $\chi^2$ into a diagonal form
\begin{equation}
  \chi_{\mathbf{S}}^{\, 2} 
  \, =  \, \alpha  \, +  \, 
  \sum_{i=1}^N \, (2 \, \beta_i \, z_i \, + \, \gamma_i \, z_i^{\,2}) 
   \label{eq:newdiag} 
\end{equation}
while preserving (\ref{eq:diagchi}), as is derived in the Appendix. 
The contribution 
$\chi_{\mathbf{\overline{S}}}^{\, 2} = \chi^2 - \chi_{\mathbf{S}}^{\, 2}$ 
from the remainder of the data $\mathbf{\overline{S}}$ is then similarly 
diagonal.

If the parameters $\gamma_i$ all lie in the range $0 < \gamma_i < 1$, 
Eqs.\ (\ref{eq:diagchi}) and (\ref{eq:newdiag}) can be written in the form
\begin{eqnarray}
  \chi^{\, 2} &=&  \chi_{\mathbf{S}}^{\, 2} \, + \, 
                   \chi_{\mathbf{\overline{S}}}^{\, 2} \nonumber \\
  \chi_{\mathbf{S}}^{\, 2}  &=&   \mathrm{const} \, + \,
        \sum_{i=1}^N \left(\frac{z_i - A_i}{B_i}\right)^2 \nonumber \\
  \chi_{\mathbf{\overline{S}}}^{\, 2} &=&  \mathrm{const} \, + \,
        \sum_{i=1}^N \left(\frac{z_i - C_i}{D_i}\right)^2 \; .
   \label{eq:BetaGamma} 
\end{eqnarray}
These equations have an obvious interpretation that is the basis of the DSD 
method: \emph{In the new coordinates, the subset $\mathbf{S}$ of the data and 
its complement $\mathbf{\overline{S}}$ take the form of independent 
measurements of the $N$ variables $z_i$} in the quadratic approximation. 
The results from Eq.\ (\ref{eq:BetaGamma}) can be read as 
\begin{eqnarray}
 z_i &=& A_i \, \pm \, B_i \; \mbox{according to } \mathbf{S} \nonumber \\
 z_i &=& C_i \, \pm \, D_i \; \mbox{according to } \mathbf{\overline{S}} 
   \label{eq:CentralResult1} 
\end{eqnarray}
where
\begin{eqnarray}
     A_i &=& -\beta_i/\gamma_i, \; B_i = 1/\sqrt{\gamma_i}   \nonumber \\
     C_i &=& \beta_i/(1-\gamma_i), \; D_i = 1/\sqrt{1-\gamma_i} \; . 
   \label{eq:CentralResult2} 
\end{eqnarray}

Eqs.\ (\ref{eq:CentralResult1})--(\ref{eq:CentralResult2})
provide a direct assessment of 
the compatibility between the subset $\mathbf{S}$ and the 
rest of the data $\mathbf{\overline{S}}\,$.  For if Gaussian statistics 
can be used to combine the uncertainties in quadrature, the difference 
between the two measurements of $z_i$ is  
\begin{eqnarray}
A_i - C_i \, &\pm& \, \sqrt{B_i^{\, 2} + D_i^{\, 2}} \nonumber \\
   = \; \frac{-\beta_i}{\gamma_i(1-\gamma_i)} \, &\pm& \,
 \frac{1}{\sqrt{\gamma_i(1-\gamma_i)}}  \; .
\end{eqnarray}
This leads to a chi-squared measure of the overall difference between 
$\mathbf{S}$ and $\mathbf{\overline{S}}$ along direction $z_i$:
\begin{equation}
   \sigma_i^{\, 2} \, = \,
   \left(\frac{A_i - C_i}{\sqrt{B_i^{\, 2} + D_i^{\, 2}}}\right)^2 \, = \, 
   \frac{\beta_i^{\, 2}}{\gamma_i \, (1 - \gamma_i)} \; .
 \label{eq:CentralMeasurements}
\end{equation}
(The symmetry of (\ref{eq:CentralMeasurements}) under the interchange 
$\gamma_i \leftrightarrow 1-\gamma_i$ reflects the obvious symmetry 
$\mathbf{S} \leftrightarrow \mathbf{\overline{S}}$.)
Even in applications where Gaussian statistics cannot be assumed, the variables 
$z_i$ are natural quantities for testing the compatibility of 
$\mathbf{S}$ with the rest of the data.

Eqs.\ (\ref{eq:CentralResult1})--(\ref{eq:CentralResult2})
also directly answer the question 
\emph{``What is measured by the subset $\mathbf{S}$ of data?''.}  
For, provided $\mathbf{S}$ is compatible with its complement, the variables 
$z_i$ that are significantly measured by $\mathbf{S}$ are 
those for which the uncertainty $B_i$ from $\mathbf{S}$ is 
less than or comparable to the uncertainty $D_i$ from $\mathbf{\overline{S}}$.

\begin{table}[htb]
  \begin{center}
  \begin{tabular}{|c|c|}
     \hline
              $\gamma_i$  &  $B_i/D_i$  \\
     \hline
                $0.9$     &  $1/3$ \\
                $0.8$     &  $1/2$ \\
                $0.5$     &  $1/1$  \\
                $0.2$     &  $2/1$  \\
                $0.1$     &  $3/1$  \\
     \hline
  \end{tabular}
\vskip -10pt
  \end{center}
  \caption{Ratio between $B_i$ = uncertainty from $\mathbf{S}$ and 
$D_i$ = uncertainty from $\mathbf{\overline{S}}$,
for various $\gamma_i\,$. 
}
  \label{table:table1}
\end{table}

For purposes of orientation, the relationship between $\gamma_i$ and the 
ratio of uncertainties $B_i/D_i = \sqrt{(1-\gamma_i)/\gamma_i}$ is shown in 
Table~\ref{table:table1} for some values of $\gamma_i$ that correspond to 
simple ratios.  In particular, $\gamma_i = 0.5$ means that $\mathbf{S}$ 
and $\mathbf{\overline{S}}$ contribute equally to the measurement of $z_i$;
while 
$\gamma_i = 0.9$ means that the uncertainty from $\mathbf{S}$ is 
three times smaller than from $\mathbf{\overline{S}}$; and 
$\gamma_i = 0.1$ means that the uncertainty from $\mathbf{S}$ is 
three times larger than from $\mathbf{\overline{S}}$.  Practically 
speaking, one can say that $\mathbf{S}$ dominates the measurement of 
$z_i$ if $\gamma_i \gtrsim 0.8-0.9$, while the complementary set 
$\mathbf{\overline{S}}$ dominates if $\gamma_i \lesssim 0.1-0.2$. 
Beyond those ranges, the contribution from the less-important 
quantity is strongly suppressed when the weighted average is taken.

Another way to interpret the $\gamma_i$ parameter is as follows.  Pretend 
that $\mathbf{S}$ consists of $N_{\displaystyle{s}}$ repeated measurements 
of $z_i$, each having the same precision; and that $\mathbf{\overline{S}}$
similarly consists of $N_{\overline{\displaystyle{s}}}$ measurements.  
The ratio of uncertainties is then given by 
\begin{equation}
 \frac{B_i}{D_i} = \sqrt{\frac{N_{\overline{\displaystyle{s}}}}
                              {N_{\displaystyle{s}}}}
\; \Longrightarrow \;
\gamma_i = \frac{N_{s}}{N_{\displaystyle{s}} + N_{\overline{\displaystyle{s}}}} 
\; .
\end{equation}
\emph{Thus $\gamma_i$ can be interpreted as the fraction of the data
that is contained in subset $\mathbf{S}$, for the purpose of measuring $z_i$.}

In applications of the DSD method, it is likely that not all of the 
$\gamma_i$ parameters will lie in the range $0 < \gamma_i < 1$.  
For if $\gamma_i \gtrsim 1$, then $\mathbf{S}$ dominates 
the measurement of $z_i$, so $\mathbf{\overline{S}}$ is 
quite insensitive to $z_i$, so the dependence of  
$\chi_{\mathbf{\overline{S}}}^{\, 2}$ on $z_i$ is likely not 
to be described well by a quadratic approximation.
Similarly $\gamma_i \lesssim 0$ means that $\mathbf{\overline{S}}$
dominates the measurement of $z_i$, so the small dependence of 
$\chi_{\mathbf{S}}^{\, 2}$ on $z_i$ may not be very quadratic. 

Compatibility between $\mathbf{S}$ and $\mathbf{\overline{S}}$ 
along directions for which $\gamma_i \gtrsim 0.8$ or 
$\gamma_i \lesssim 0.2$ is not a crucial issue, since one or the 
other measurement dominates the average along such directions.
\emph{It is an important feature of the DSD method that it 
distinguishes between inconsistencies that do or do not affect 
the overall fit.}  In that sense, it is a more sensitive tool than 
the previous method of simply studying $\chi_{\mathbf{S}}^{\, 2}$ 
vs. $\chi_{\mathbf{\overline{S}}}^{\, 2}$ by means of a variable 
weight \cite{Collins}.

\section{Applications to parton distribution analysis}
\label{sec:Applications}

The interpretation of data from high energy colliders such as the Tevatron 
at Fermilab and the LHC at CERN relies on knowing the PDFs that describe 
momentum distributions of quarks and gluons in the proton.  These PDFs are 
extracted by a ``global analysis'' \cite{CT09,MSTW08} of many kinds of 
experiments whose results are tied together by the theory of Quantum 
Chromodynamics (QCD).  The analysis described here to illustrate the DSD method 
is based on 36 data sets with a total of 2959 data points. These are the same 
data sets used in a recent PDF analysis \cite{CT09}, except that two older 
inclusive jet experiments have been dropped for simplicity.

The theory uses the same 24 free parameters as that recent analysis.  
These parameters describe the momentum distributions 
$u(x)$, $d(x)$, $\bar{u}(x)$, $\bar{d}(x)$, $\bar{s}(x)$ and $g(x)$ at 
a particular small QCD scale. All of the PDFs at higher scale can 
be calculated from these by QCD.

This PDF application is a strong test of the new method, because 
the large number of experiments of different types carries the possibility 
for unknown experimental and theoretical systematic errors, and the large 
number of free parameters includes a wide range of flat and steep directions 
in parameter space.

\subsection{E605 experiment} 
We first apply the data set diagonalization method to study the contribution 
of the E605 experiment \cite{E605} to the PDF analysis.  This experiment 
(lepton pair production in proton scattering on copper) is sensitive 
to the various flavors of quarks in the proton in a different way from the 
majority of the data, so it can be expected to be responsible for one or more 
specific features of the global fit.  It is also an experiment where unknown 
systematic errors might be present, since no corrections for possible nuclear 
target effects are included.

There are 24 free parameters in the fit, and hence 24 mutually 
orthogonal eigenvector directions.  In descending order, the 
first 4 of these are found to have 
$\gamma_1 = 0.91$, 
$\gamma_2 = 0.38$, 
$\gamma_3 = 0.16$, 
$\gamma_4 = 0.06\,$.  
All of the other eigenvectors have still smaller or even negative $\gamma_i$.  
Hence according to the previous discussion,
the fit is controlled mainly by this E605 data set along eigenvector 
direction 1; E605 and its complement both play a role along 
direction 2; E605 plays a very minor role along direction 3; and it is 
unimportant along the remaining 21 directions. 

% f474U.x >& ahhai.con hx12
%cvec(1)= 0.90703E+00 fsub -0.3710 +/- 1.0728 fcomp  2.9393 +/- 2.6679 diff= -3.3103 +/- 2.8755 (1.1512 sigma)
%cvec(2)= 0.38068E+00 fsub -1.3848 +/- 1.6131 fcomp  0.8707 +/- 1.2935 diff= -2.2555 +/- 2.0677 (1.0908 sigma)
%cvec(3)= 0.16387E+00 fsub  0.0518 +/- 2.4501 fcomp -0.0103 +/- 1.0966 diff=  0.0621 +/- 2.6843 (0.0231 sigma)
%cvec(4)= 0.63606E-01 fsub  1.5704 +/- 3.9152 fcomp -0.1015 +/- 1.0311 diff=  1.6719 +/- 4.0487 (0.4129 sigma)
% ipar  cvec(ipar)         fsub                complement              difference         sigma   sigma**2
%   1  0.9070E+00   -0.363 +/-    1.050     3.541 +/-    3.280     3.904 +/-    3.444     1.134     1.29
%   2  0.3807E+00   -1.390 +/-    1.621     0.854 +/-    1.271     2.245 +/-    2.060     1.090     1.19
%   3  0.1639E+00    0.052 +/-    2.470    -0.010 +/-    1.094     0.063 +/-    2.702     0.023     0.00
%   4  0.6361E-01    1.498 +/-    3.965    -0.102 +/-    1.033     1.599 +/-    4.098     0.390     0.15

\begin{figure}
\begin{center}
 \resizebox*{0.48\textwidth}{!}{
\includegraphics[clip=true,scale=1.0]{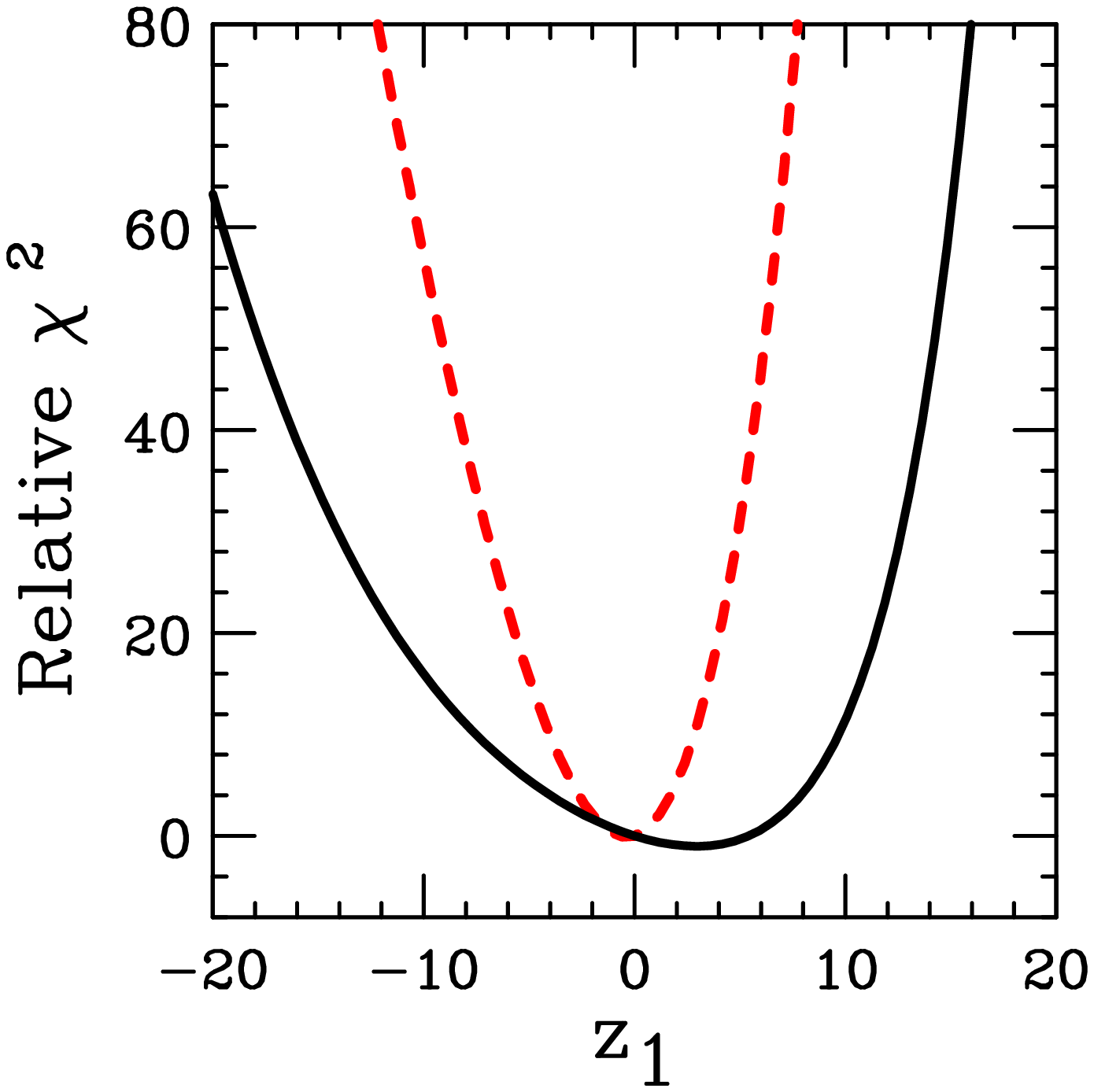}
\hfill
\includegraphics[clip=true,scale=1.0]{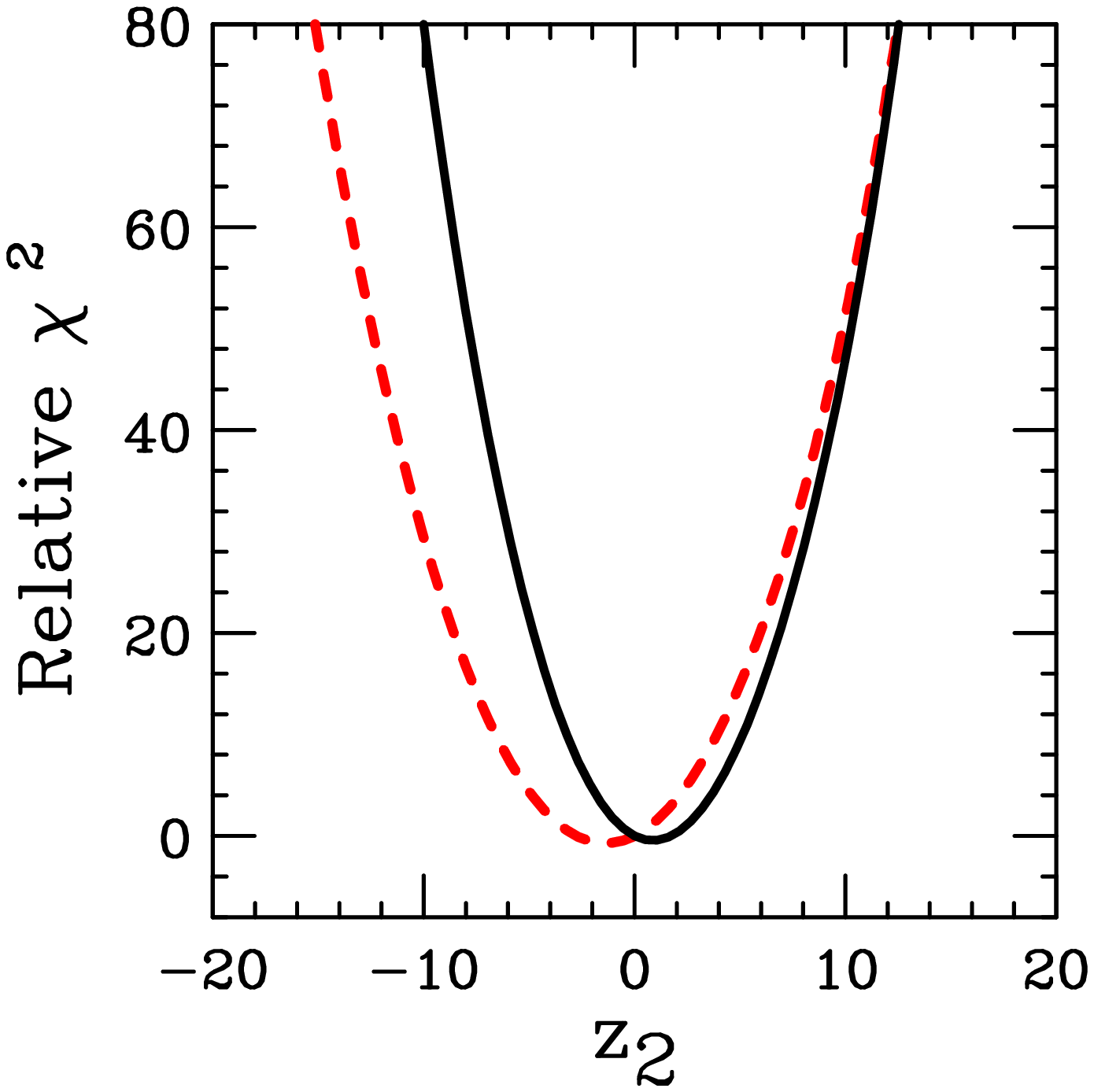}}
\end{center}
\begin{center}
 \resizebox*{0.48\textwidth}{!}{
\includegraphics[clip=true,scale=1.0]{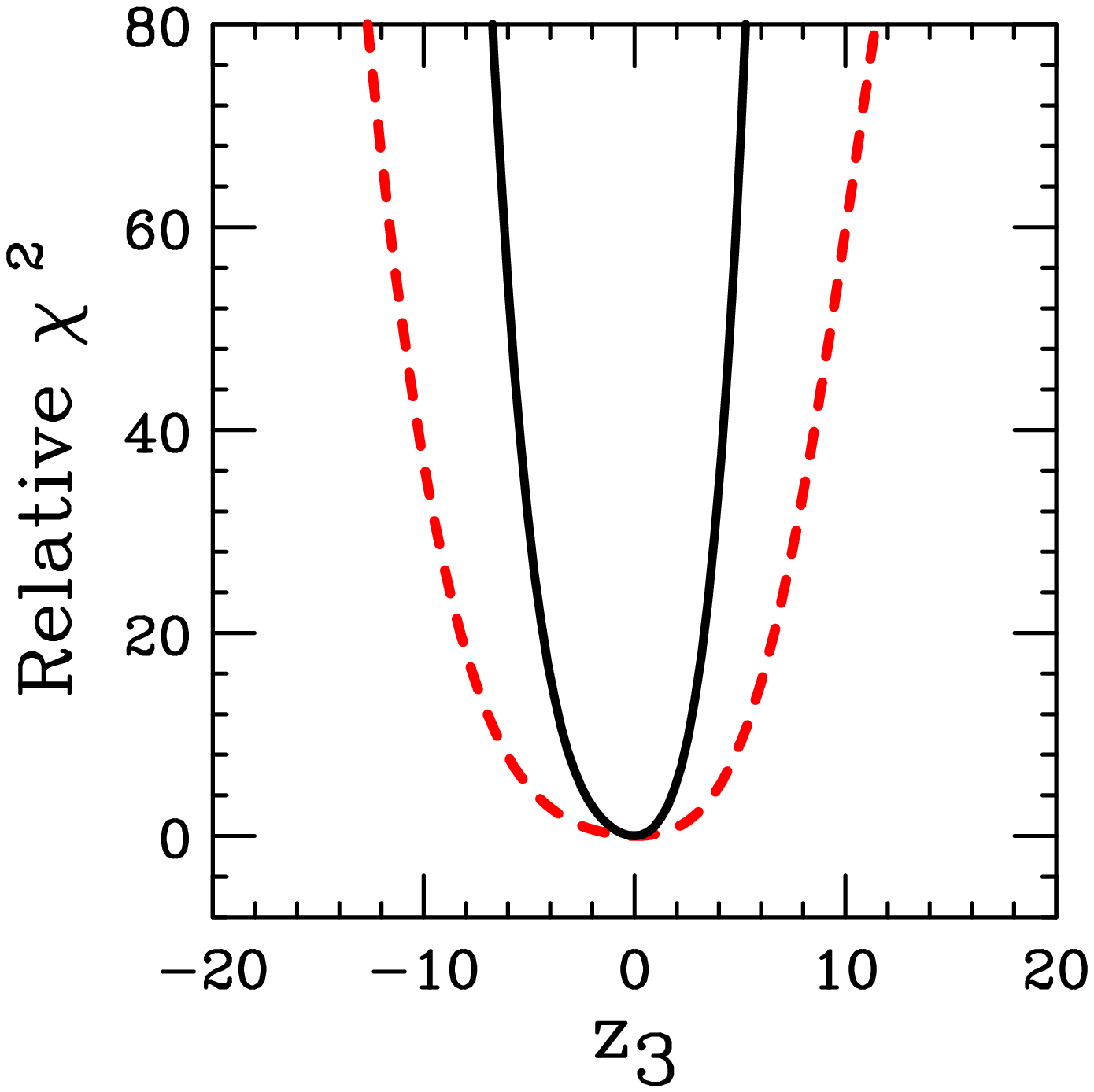}
\hfill
\includegraphics[clip=true,scale=1.0]{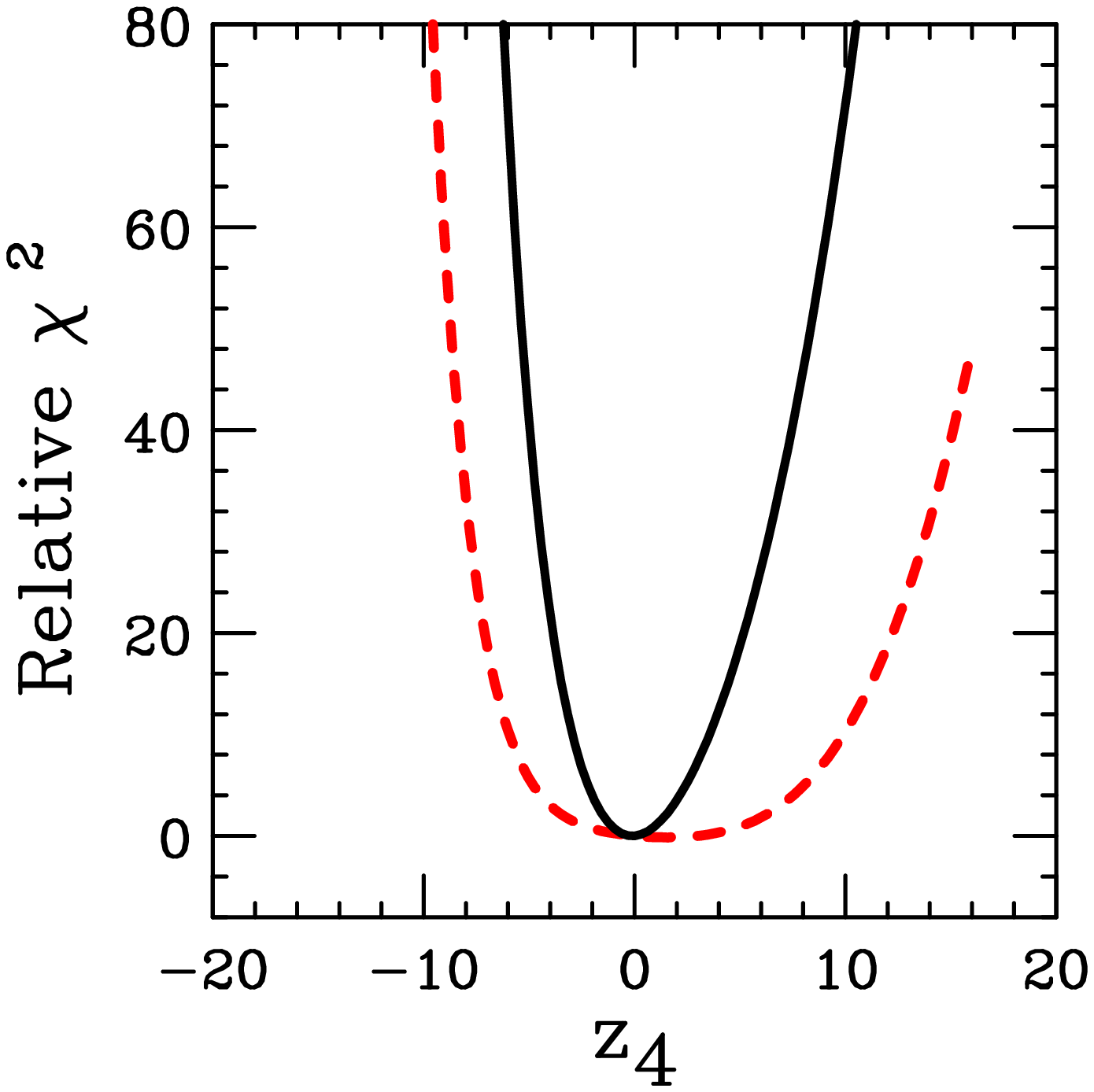}}
\end{center}
 \caption{$\chi^2$ for fit to E605 (dashed curves) and to the 
rest of the data (solid curves) along the four leading eigenvector 
directions in descending order of $\gamma_i$.  In each panel, $z_i = 0$ 
is the location of the overall best fit.}
 \label{fig:figOne}
\end{figure}

This is confirmed in Fig.~\ref{fig:figOne}, which shows the variation 
of $\chi^2$, with the best-fit values subtracted, for E605 (119 data points) 
and its complement (the remaining 2840 data points) along each of the first 
four directions.  Along direction 1, the E605 data indeed dominate the 
measurement:  the ``parabola'' of $\chi^2_{\mathbf{S}}$ is much narrower than 
the ``parabola'' of $\chi^2_{\mathbf{\overline{S}}}$.  The minimum for the 
complementary data set $\mathbf{\overline{S}}$ lies rather far from the best 
fit value $z_1 = 0$, but its $\chi^2$ is so slowly varying that it is not 
inconsistent with that value.
Along direction 2, E605 and its complement are both important, and the two 
measures are again seen to be consistent with each other.  For the 
remaining 2 directions shown, and the 20 directions that are not shown, 
the $\mathbf{\overline{S}}$ data completely dominate: E605 provides negligible 
information along those directions.
(The $z_4$ curve for E605 ends abruptly, because the fit becomes numerically 
unphysical at that point, which is far outside the region of acceptable fits 
to $\mathbf{\overline{S}}$.)

\begin{table}[htb]
  \begin{center}
\null
\begin{tabular}{||c|c||c|c||c|c||}
\hline
$i$ & $\gamma_i$  & $z_i$ from $\mathbf{S}$ & $z_i$ from $\mathbf{\overline{S}}$  & 
Difference & $\sigma_i$ \\
\hline
%
% version 1 version of table 2 -- based on weighted chisqr for the non-E605 data
% $1$ & $0.93$ & $      -0.40 \pm 1.06$ & $\quad 3.78 \pm 2.55$ & $4.18 \pm 2.76$ & $1.51 \,$  \\
% $2$ & $0.42$ & $      -1.22 \pm 1.52$ & $\quad 0.93 \pm 1.34$ & $2.15 \pm 2.03$ & $1.06 \,$  \\
% $3$ & $0.09$ & $ \quad 1.71 \pm 3.34$ & $     -0.15 \pm 1.01$ & $1.86 \pm 3.48$ & $0.53 \,$  \\
% $4$ & $0.05$ & $      -0.57 \pm 4.54$ & $\quad 0.03 \pm 1.03$ & $0.60 \pm 4.65$ & $0.13 \,$  \\
%
% f474U.x >& ahhai.con hx12
% version 2 version of table 2 -- based on fits where all experiments have weight 1
 $1$&$0.91$&$     -0.37 \pm 1.07$&$\quad 2.94 \pm 2.67$&$     -3.31 \pm 2.88$&$1.15 \,$ \\
 $2$&$0.38$&$     -1.38 \pm 1.61$&$\quad 0.87 \pm 1.29$&$     -2.26 \pm 2.07$&$1.09 \,$ \\
 $3$&$0.16$&$\quad 0.05 \pm 2.45$&$     -0.01 \pm 1.10$&$\quad 0.06 \pm 2.68$&$0.02 \,$ \\
 $4$&$0.06$&$\quad 1.57 \pm 3.92$&$     -0.10 \pm 1.03$&$\quad 1.67 \pm 4.05$&$0.41 \,$ \\
%
%cvec(1)= 0.90703E+00 fsub -0.3710 +/- 1.0728 fcomp  2.9393 +/- 2.6679 diff= -3.3103 +/- 2.8755 ( 1.1512 sigma)
%cvec(2)= 0.38068E+00 fsub -1.3848 +/- 1.6131 fcomp  0.8707 +/- 1.2935 diff= -2.2555 +/- 2.0677 ( 1.0908 sigma)
%cvec(3)= 0.16387E+00 fsub  0.0518 +/- 2.4501 fcomp -0.0103 +/- 1.0966 diff=  0.0621 +/- 2.6843 ( 0.0231 sigma)
%cvec(4)= 0.63606E-01 fsub  1.5704 +/- 3.9152 fcomp -0.1015 +/- 1.0311 diff=  1.6719 +/- 4.0487 ( 0.4129 sigma)
\hline
\end{tabular}
\vskip -10pt
  \end{center}
  \caption{Consistency beween $\mathbf{S}$ = E605 experiment and 
$\mathbf{\overline{S}}$ = the remainder of data.}
  \label{table:table2}
\end{table}
The $\mathbf{S}$ and $\mathbf{\overline{S}}$ columns of 
Table~\ref{table:table2} show the information of Fig.~\ref{fig:figOne} 
interpreted as measurements of $z_1,\dots,z_4\,$.  This can be done 
according to Eqs.\ (\ref{eq:BetaGamma})--(\ref{eq:CentralResult2}), or more 
precisely by fitting each of the curves in Fig.~\ref{fig:figOne} 
to a parabolic form in the neighborhood of its minimum rather than fitting 
at $z_i = 0$.
The Difference column is the difference between the 
$\mathbf{S}$ and $\mathbf{\overline{S}}$ measurements of $z_i$, with an 
error estimate obtained by adding the $\mathbf{S}$ and $\mathbf{\overline{S}}$ 
errors in quadrature.  The final column expresses this difference in units 
of its uncertainty, which would be the number of standard deviations for 
Gaussian statistics.  The fact that these numbers are $\lesssim 1$
implies that the E605 experiment is consistent with the rest of the 
global analysis.

\subsection{Inclusive jet experiments} 
We now turn our attention to the role of the CDF \cite{CDFR2} and 
D0 \cite{D0R2} run II jet experiments in the PDF analysis.  This was the 
principal subject of a recent paper \cite{CT09}; but the DSD technique can 
shed new light on it.  We first examine the consistency between each jet 
experiment and the rest of the data with the other jet experiment excluded.  
Results for the leading $\gamma_i$ are shown in Table~\ref{table:table3} 
for CDF and Table~\ref{table:table4} for D0.  The CDF experiment plays a 
strong role along its two leading directions 
($\gamma_1=0.75$ and $\gamma_2=0.62$), showing a rather strong tension
($3.6 \, \sigma$) along $z_2$.
The D0 experiment similarly plays a strong role along its two leading 
directions ($\gamma_1 = 0.71$ and $\gamma_2 = 0.52$), but it is consistent 
with the non-jet data along both of those directions.

\begin{table}[htb]
  \begin{center}
\begin{tabular}{||c|c||c|c||c|c||}
\hline
$i$ & $\gamma_i$  & $z_i$ from $\mathbf{S}$ & $z_i$ from $\mathbf{\overline{S}}$  & 
Difference & $\sigma_i$ \\
\hline
%CDF run II jets only f474U.x >& ahhyr.con hx7 
 $1$ & $0.75$ & $  0.55 \pm 1.11$ & $-1.74 \pm 1.85$ & $  2.28 \pm 2.15$ & $1.06 \,$ \\
 $2$ & $0.62$ & $  2.66 \pm 1.25$ & $-4.34 \pm 1.52$ & $  7.00 \pm 1.96$ & $3.56 \,$ \\
 $3$ & $0.04$ & $ 11.26 \pm 4.14$ & $-0.58 \pm 1.03$ & $ 11.84 \pm 4.26$ & $2.78 \,$ \\
\hline
\end{tabular}
\vskip -10pt
  \end{center}
  \caption{Consistency between $\mathbf{S}$ = CDF and 
           $\mathbf{\overline{S}}$ = all non-jet data.}
  \label{table:table3}
\end{table}

\begin{table}[htb]
  \begin{center}
\begin{tabular}{||c|c||c|c||c|c||}
\hline
$i$ & $\gamma_i$  & $z_i$ from $\mathbf{S}$ & $z_i$ from $\mathbf{\overline{S}}$  & 
Difference & $\sigma_i$ \\
\hline
%D0 run II jets only f474U.x >& ahhzq.con hx10 
 $1$ & $0.71$ & $\quad  0.49 \pm 1.11$ & $     -1.33 \pm 1.79$ & $\quad  1.82 \pm 2.11$ & $0.86 \,$ \\
 $2$ & $0.52$ & $\quad  1.05 \pm 1.36$ & $     -1.26 \pm 1.51$ & $\quad  2.31 \pm 2.03$ & $1.14 \,$ \\
 $3$ & $0.07$ & $      -2.00 \pm 3.89$ & $\quad 0.14 \pm 1.03$ & $      -2.14 \pm 4.02$ & $0.53 \,$ \\
\hline
\end{tabular}
\vskip -10pt
  \end{center}
  \caption{Consistency between $\mathbf{S}$ = D0 and 
           $\mathbf{\overline{S}}$ = all non-jet data.}
  \label{table:table4}
\end{table}

Since these jet experiments measure the same process by similar techniques, 
it also makes sense to combine them into a single subset $\mathbf{S}$.  
The result is given in Table~\ref{table:table5}.  The $\gamma_i$ parameters 
in descending order are 
$\gamma_1 = 0.82$, $\gamma_2 = 0.74$, $\gamma_3 = 0.12$, $\gamma_4 = 0.05$,  
so these data supply most of the constraint along their two leading
directions, and negligible constraint along any of the others.  
The expectation that these two experiments measure the same thing is confirmed 
by the fact that there are still only two directions being determined, with 
$\gamma_1$ and $\gamma_2$ larger than for either experiment alone.  
Some tension ($2.8 \, \sigma$) exists between $\mathbf{S}$ and 
$\mathbf{\overline{S}}$ along $z_2$; but combining the data sets has reduced 
the conflict relative to what appeared with CDF alone.

\begin{table}[htb]
  \begin{center}
\begin{tabular}{||c|c||r|r||r|r||}
\hline
$i$ & $\gamma_i$  & $z_i$ from $\mathbf{S}$ & $z_i$ from $\mathbf{\overline{S}}$  & 
Difference & $\sigma_i$ \\
\hline
%fsub=CDF+D0 (run II only) f474U.x >& ahhxr.con
%cvec( 1)= 0.81737E+00 fsub  0.3493 +/- 1.0786 fcomp -1.6756 +/- 2.3088 diff=  2.0249 +/-  2.5483 (0.7946 sigma)
%cvec( 2)= 0.73618E+00 fsub  1.6249 +/- 1.1480 fcomp -4.6041 +/- 1.8921 diff=  6.2290 +/-  2.2131 (2.8146 sigma)
%cvec( 3)= 0.12490E+00 fsub -0.1854 +/- 2.8414 fcomp  0.0263 +/- 1.0678 diff= -0.2117 +/-  3.0354 (0.0697 sigma)
%cvec( 4)= 0.50926E-01 fsub  3.1445 +/- 4.3368 fcomp -0.1627 +/- 0.9727 diff=  3.3072 +/-  4.4446 (0.7441 sigma)
%still only two directions -- now better defined -- because the two expts measure the same thing
 $1$ & $0.82$ & $      0.35 \pm 1.08$ & $     -1.68 \pm 2.31$ & $ 2.02 \pm  2.55$ & $0.79 \,$ \\
 $2$ & $0.74$ & $      1.62 \pm 1.15$ & $     -4.60 \pm 1.89$ & $ 6.23 \pm  2.21$ & $2.81 \,$ \\
 $3$ & $0.12$ & $     -0.19 \pm 2.84$ & $      0.03 \pm 1.07$ & $-0.21 \pm  3.04$ & $0.07 \,$ \\
 $4$ & $0.05$ & $      3.14 \pm 4.34$ & $     -0.16 \pm 0.97$ & $ 3.31 \pm  4.44$ & $0.74 \,$ \\
\hline
\end{tabular}
\vskip -10pt
  \end{center}
  \caption{Consistency between $\mathbf{S}$ = CDF + D0 jet data and 
           $\mathbf{\overline{S}}$ = all non-jet data.}
  \label{table:table5}
\end{table}

\begin{figure}[htb]
\begin{center}
 \resizebox*{0.48\textwidth}{!}{
\includegraphics[clip=true,scale=1.0]{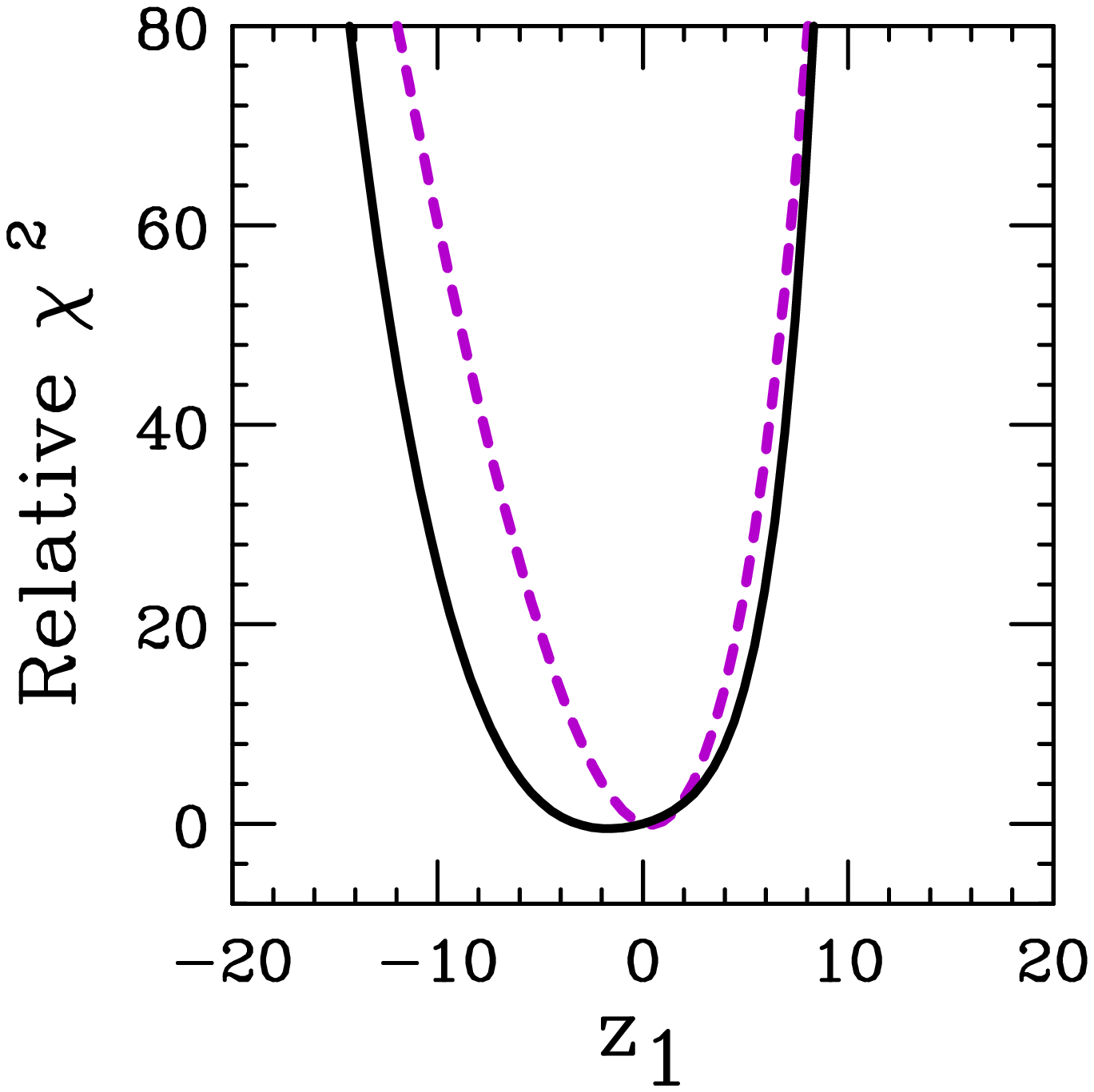}
\hfill
\includegraphics[clip=true,scale=1.0]{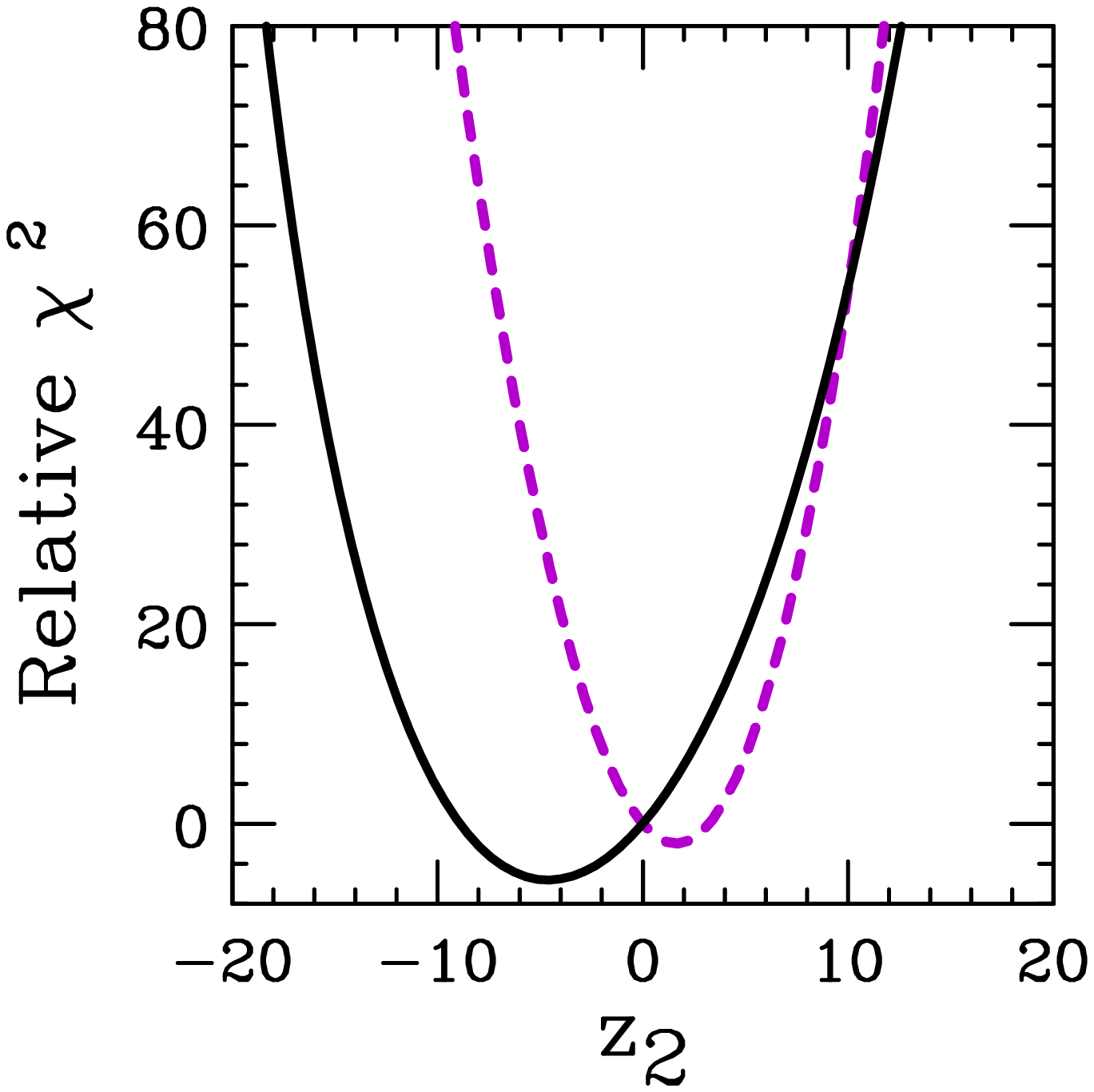}}
\end{center}
\begin{center}
 \resizebox*{0.48\textwidth}{!}{
\includegraphics[clip=true,scale=1.0]{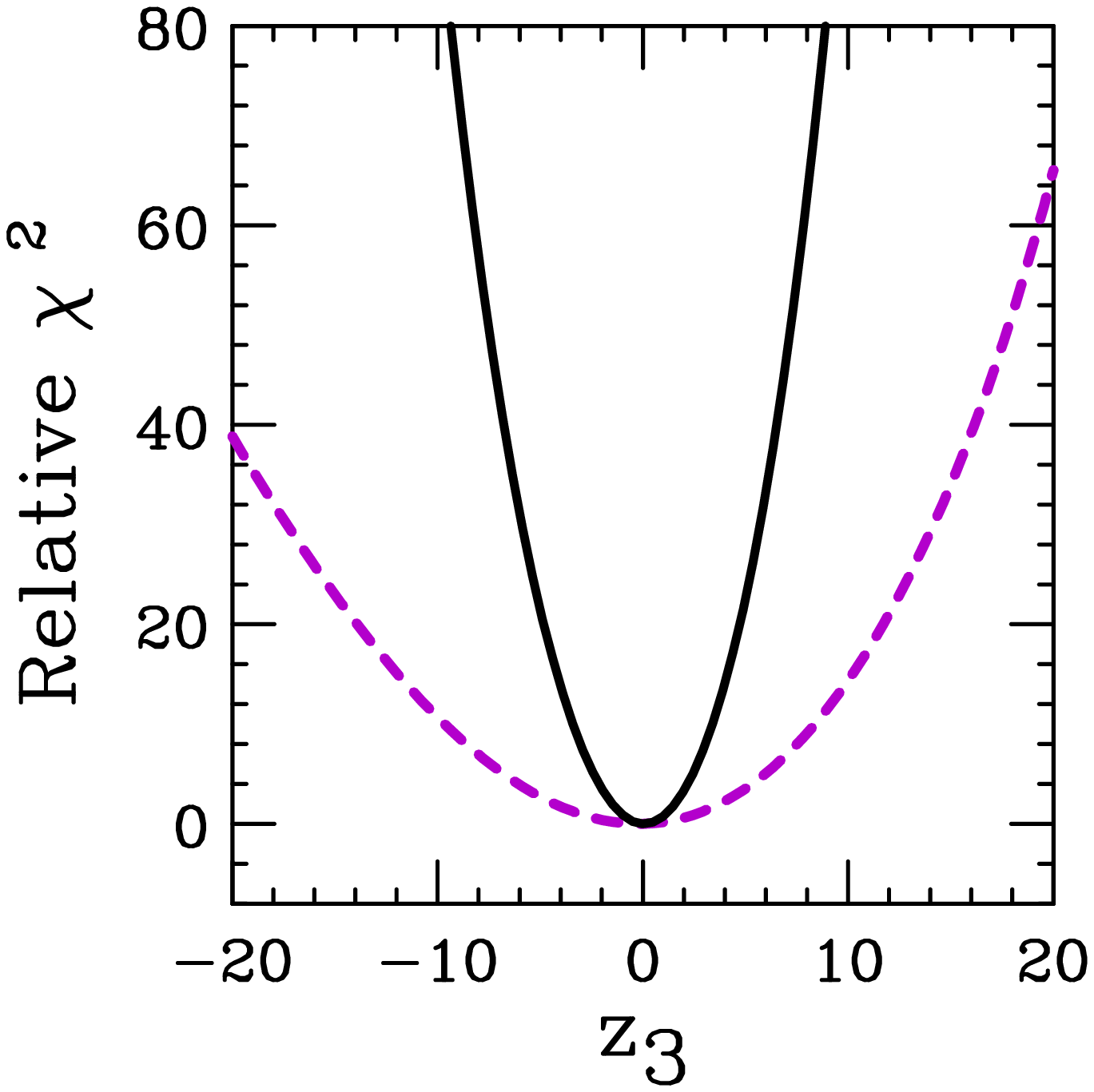}
\hfill
\includegraphics[clip=true,scale=1.0]{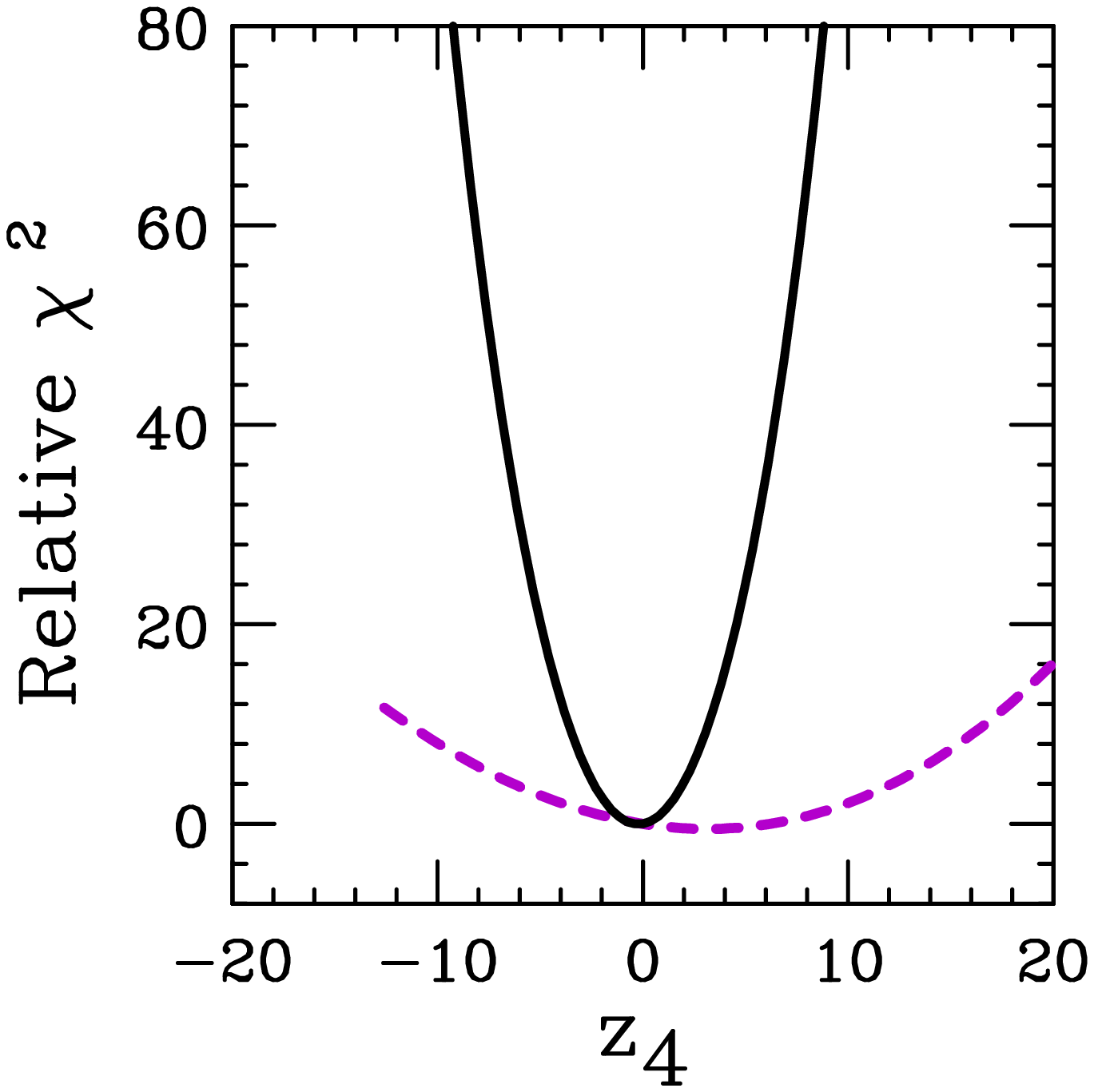}}
\end{center}
 \caption{
$\chi^2$ for fit to CDF+D0 (dashed curves) and
to the remaining data (solid curves),
for the four leading directions in descending order of $\gamma_i$.}
 \label{fig:figTwo}
\end{figure}

Figure \ref{fig:figTwo} shows the variation in $\chi^2$ for the fit to the jet 
data (72 + 110 points) and its complement (2777 points) along the four leading
directions. The numerical results shown in Table~\ref{table:table5} correspond 
to fitting these curves by parabolas at their minima.  
For the first two directions, the ``parabola'' for the jet data
$\mathbf{S}$ is narrower than the ``parabola'' for its complement, as expected 
since $\gamma_1, \gamma_2 > 0.5\,$. This confirms that the jet data dominate 
the global fit along those directions.
For $z_3$ and $z_4$ (and all other directions, which are not shown), the jet 
data supply very little constraint: the $\chi^2$ ``parabola'' is much 
broader for $\mathbf{S}$ than for $\mathbf{\overline{S}}$.
The locations of the minima are quite far apart for $z_2$, which reflects the 
tension between $\mathbf{S}$ and $\mathbf{\overline{S}}$ along that direction.  

\begin{figure}[htb]
\begin{center}
 \resizebox*{0.48\textwidth}{!}{
\includegraphics[clip=true,scale=1.0]{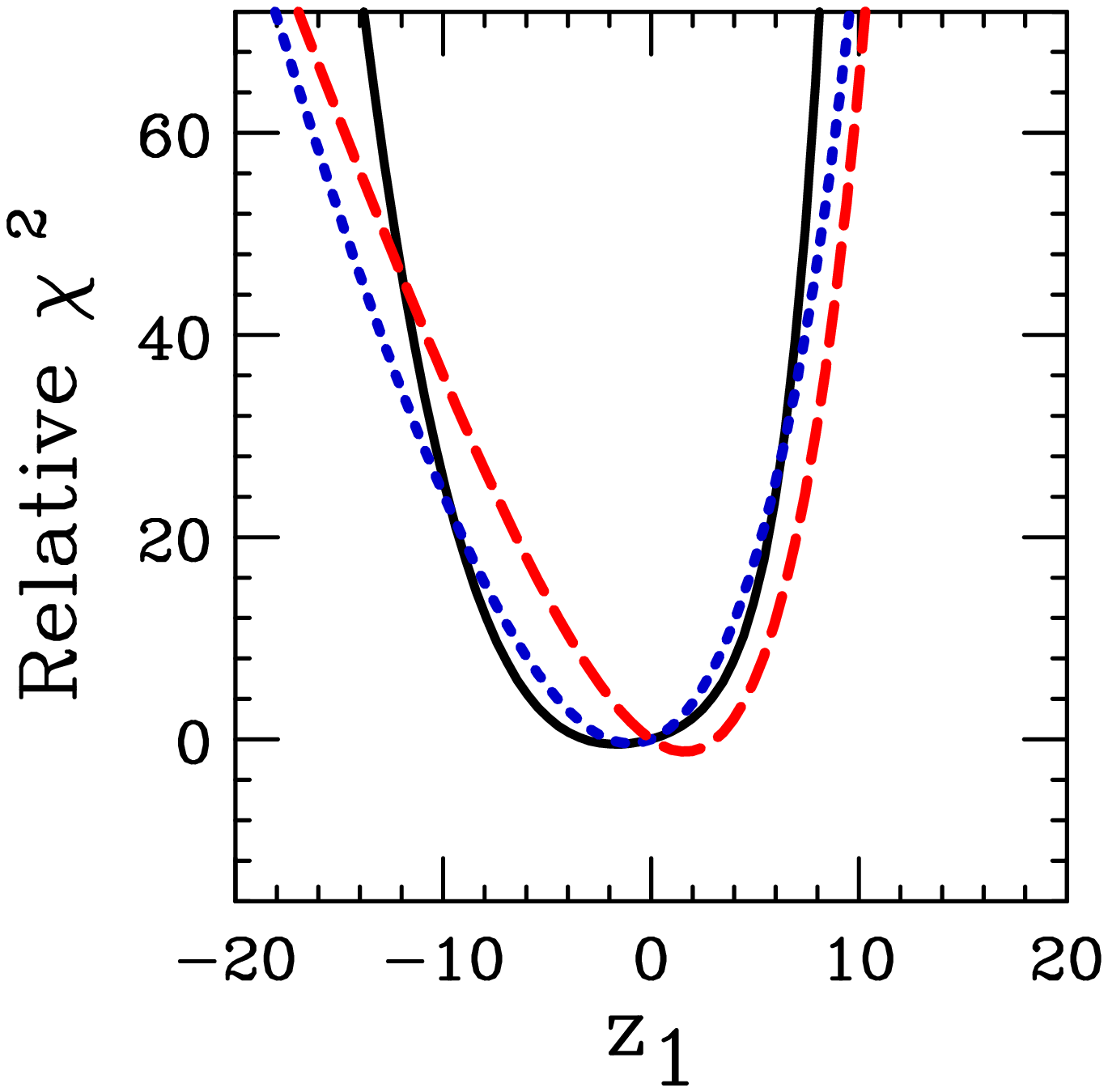}
\hfill
\includegraphics[clip=true,scale=1.0]{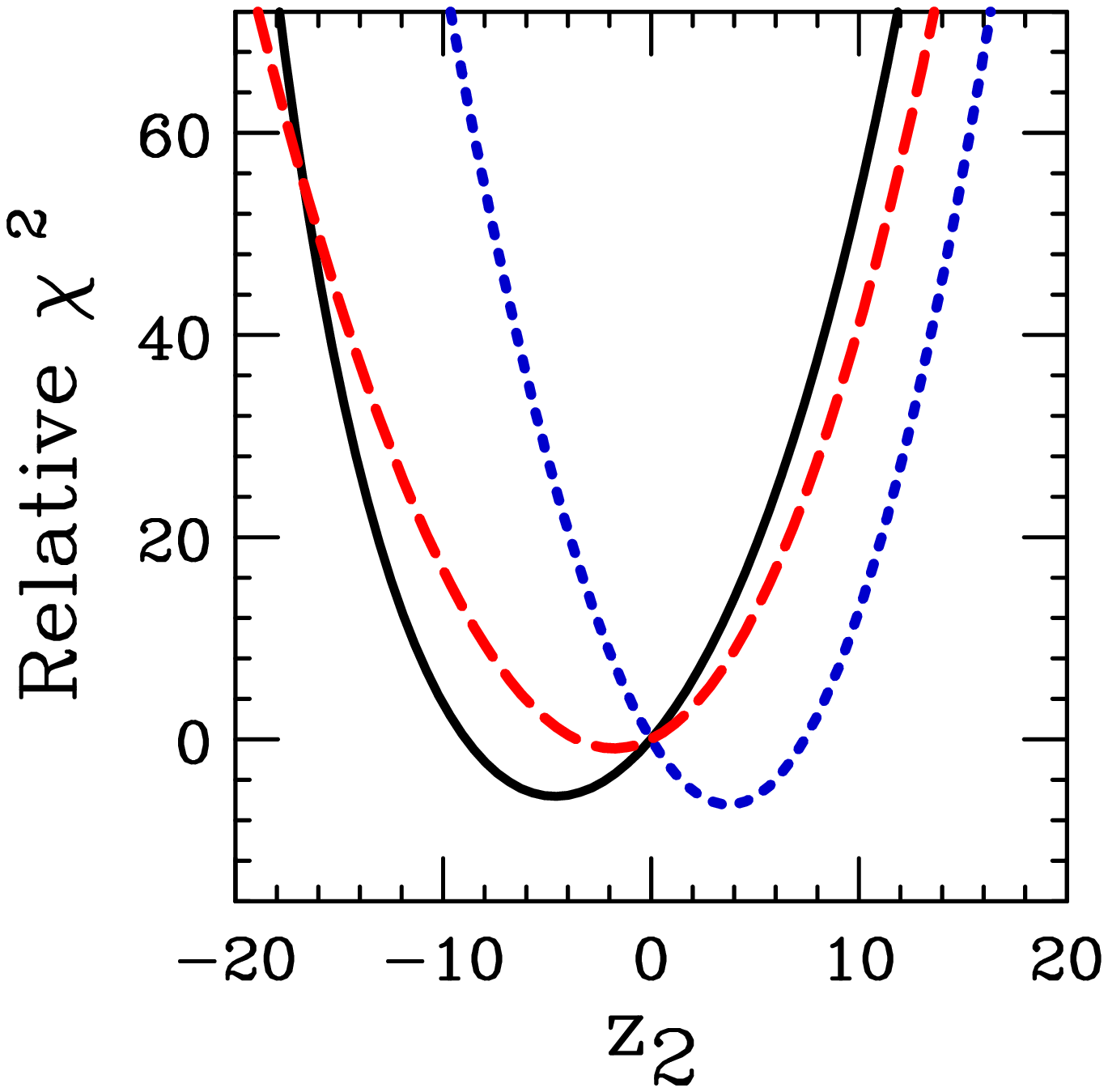}}
\end{center}
\begin{center}
 \resizebox*{0.48\textwidth}{!}{
\includegraphics[clip=true,scale=1.0]{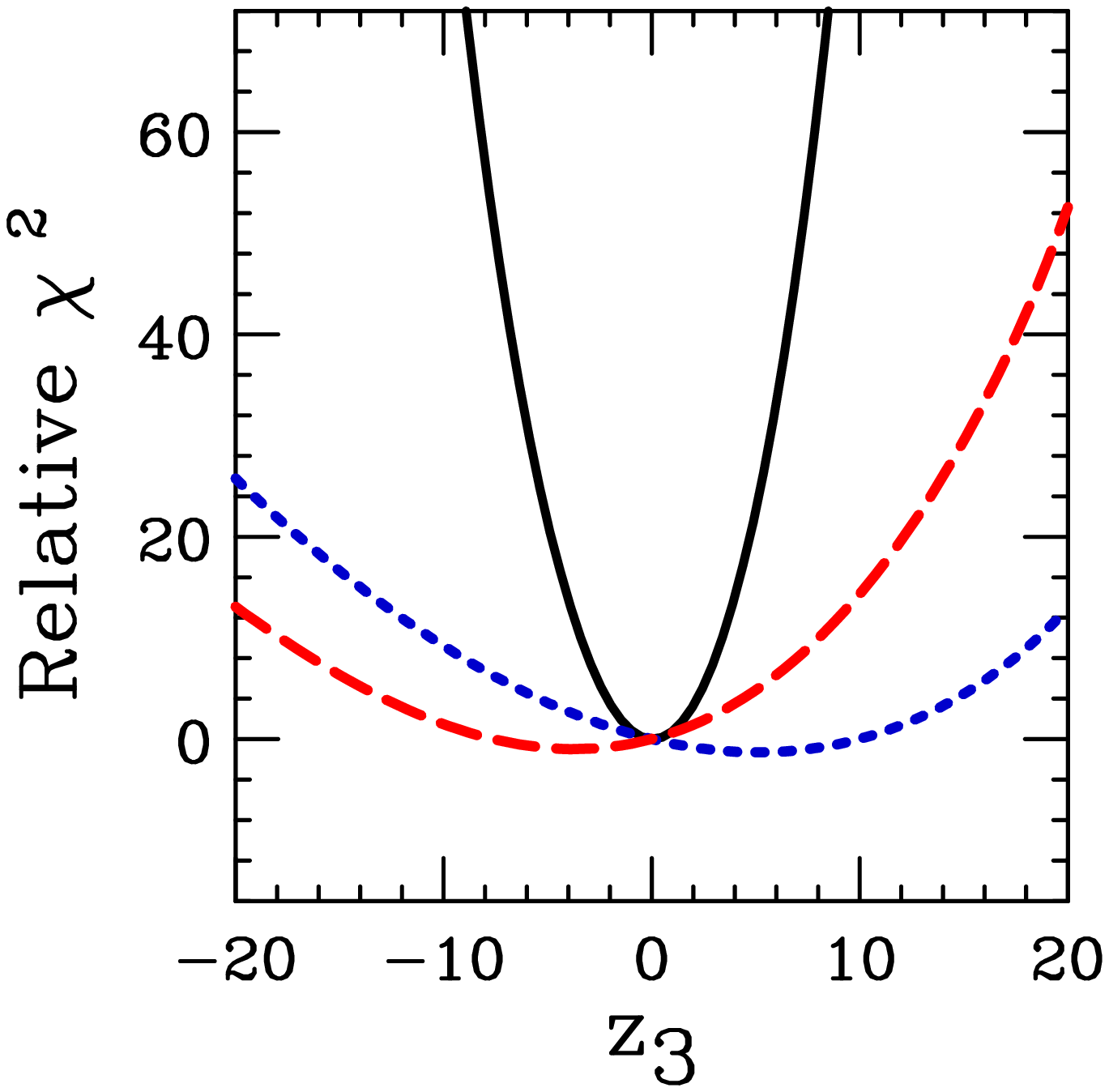}
\hfill
\includegraphics[clip=true,scale=1.0]{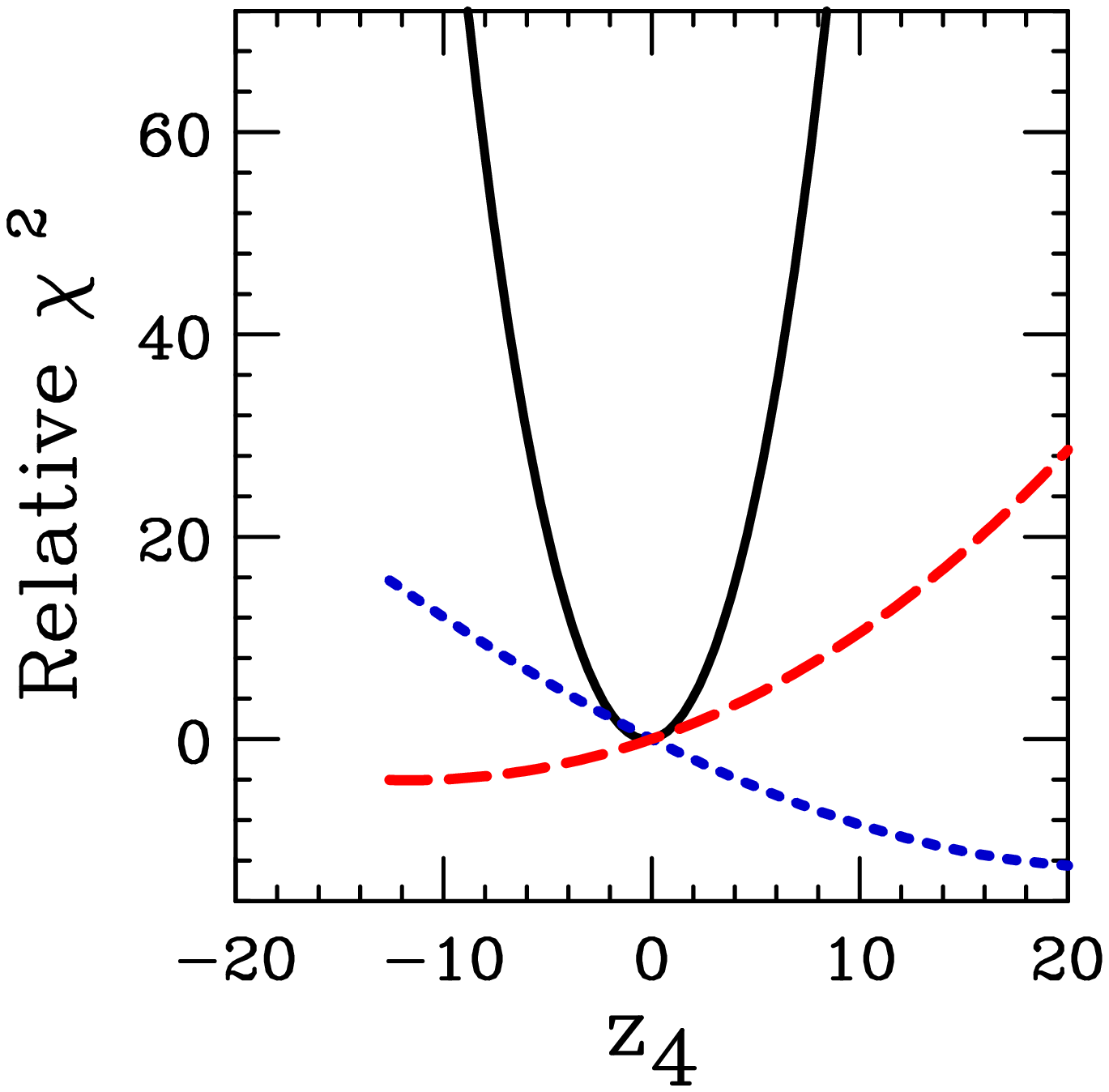}}
\end{center}
 \caption{
$\chi^2$ for fit to CDF (dotted), D0 (dashed), and 
the rest of the data (solid).
}
 \label{fig:figThree}
\end{figure}

To study the consistency between the two individual jet experiments within the 
context of the global fit, their $\chi^2$ values are plotted separately in 
Fig.~\ref{fig:figThree} along the same eigenvector directions 
as in Fig.~\ref{fig:figTwo}.
There appears to be a bit of tension between the two experiments along 
these directions, since their minima occur at different 
places. Quantitatively, fitting each curve in Fig.~\ref{fig:figThree} to a 
parabola near its minimum, leads to the results shown in 
Table~\ref{table:table6}.  The discrepancy between the jet 
experiments is $2.4 \, \sigma$ and $1.6 \, \sigma$ along the two directions 
in which these experiments are significant in the global fit.
Any discrepancy between the jet experiments along other directions, including 
the strong difference along direction 4, is not important for the global fit, 
because non-jet experiments supply much stronger constraints along those 
directions, as is confirmed by the narrow parabola for $\mathbf{\overline{S}}$.
\begin{table}[htb]
  \begin{center}
\begin{tabular}{||c||c|c||c|c||}
\hline
$i$ & $z_i$ from CDF & $z_i$ from D0 & 
Difference & $\sigma_i$ \\
\hline
 $1$ & $ 2.70 \pm 1.65$ & $-2.45 \pm 1.38$ & $ 5.15 \pm 2.15$ & $2.40 \,$ \\ 
 $2$ & $ 2.33 \pm 1.35$ & $-1.74 \pm 2.22$ & $ 4.07 \pm 2.60$ & $1.57 \,$ \\ 
\hline
\end{tabular}
\vskip -10pt
  \end{center}
  \caption{Consistency between CDF and D0 jet experiments.}
  \label{table:table6}
\end{table}

\begin{figure}[htb]
\begin{center}
 \resizebox*{0.45\textwidth}{!}{
\includegraphics[clip=true,scale=1.0]{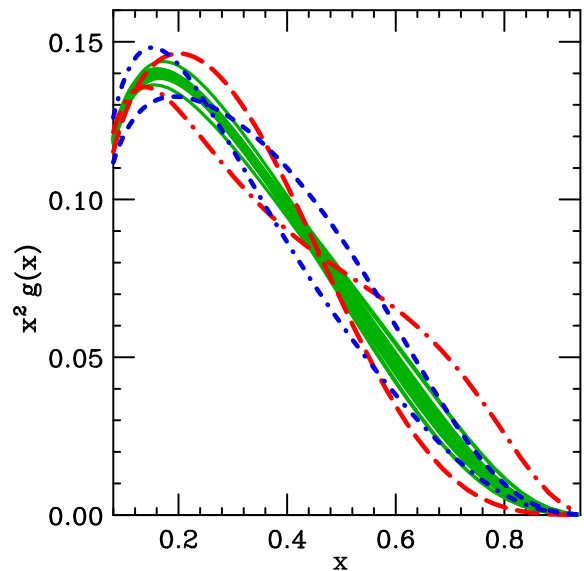}}
\end{center}
 \caption{
Gluon distributions $g(x)$ at 
$z_1 = 4.0$ (long dash), 
$z_1 = -4.0$ (long dash dot), 
$z_2 = 4.0$ (short dash), 
$z_2 = -4.0$ (short dash dot), and
$z_i = \pm \, 4.0$ for $i=3,\dots,24$ (solid).
Most of the uncertainty for $g(x)$ comes from eigenvector 
directions 1 and 2, which are controlled principally by the jet 
experiments according to Fig.\ \ref{fig:figTwo}.
}
 \label{fig:figFour}
\end{figure}

The DSD method can also be used to discover which aspects of a global fit are 
determined by particular subsets of the data.  An example of this is 
illustrated by Fig.\ \ref{fig:figFour}, which shows the gluon distribution 
at QCD scale $1.3 \, \mathrm{GeV}$, for PDF sets corresponding to displacements 
$z_i = \pm 4$ along each eigenvector direction of the CDF+D0 fit.  Most of the 
uncertainty is seen to come from the $z_1$ and $z_2$ directions, which are the 
directions found above to be controlled by the jet data.  This directly 
confirms the conclusion of \cite{CT09} that the jet data are the major source 
of information about the gluon distribution for $x \gtrsim 0.1$.

\section{Conclusion}
\label{sec:conclusion}

A ``data set diagonalization'' (DSD) procedure has been presented, which 
extends the Hessian method \cite{HessianMethod} for uncertainty analysis.  
The procedure identifies 
the directions in parameter space along which a given subset $\mathbf{S}$ of 
data provides significant constraints in a global fit. This allows one to test 
the consistency between $\mathbf{S}$ and the remainder of the data, 
and to discover which aspects of the fit are controlled by $\mathbf{S}$.

The procedure involves ``rediagonalizing'' $\chi^2$ to obtain a new set 
of fitting parameters $\{z_i\}$ that are linear combinations of the original 
ones.  The data from a given experiment or other chosen subset $\mathbf{S}$ 
of the data and its complement $\mathbf{\overline{S}}$ take the form of 
independent measurements of these new parameters, within the scope of the 
quadratic approximation.  The degree of consistency between 
$\mathbf{S}$ and $\mathbf{\overline{S}}$ can thus be examined by 
standard statistical methods.

The DSD method can be used to study the internal consistency
of a global fit, by applying it with $\mathbf{S}$ defined by each experimental 
data set in turn. One can also let $\mathbf{S}$ correspond to subsets of 
the data that are suspected of being subject to some particular kind of 
unquantified systematic error.  
A full systematic study of the parton distribution fit using the new 
technique is currently in progress.

Typical applications of the new technique have been illustrated in the 
context of measuring parton distribution functions.  The method uncovered 
and quantified tension between the two inclusive jet experiments, and 
between one of those experiments and the non-jet data, that was
difficult to detect using the older methods, which are based on tracking the 
effect on $\chi^2$ for $\mathbf{S}$ and $\mathbf{\overline{S}}$ in response 
to changing the weight assigned to $\mathbf{S}$ \cite{CT09,Collins}.

The DSD method can be also be used to identify which features of 
the fit are controlled by particular experiments or other subsets of the data 
in a complex data set.  As an example of this, the jet experiments were shown
to be the principal source of information on the gluon distribution 
in the region displayed in Fig.\ \ref{fig:figFour}.  The logic is as follows: 
Fig.\ \ref{fig:figFour} shows that the uncertainty of the gluon distribution 
is dominated by eigenvector directions 1 and 2 when $\mathbf{S}$ is defined 
as the jet data; and the range of acceptable fits along those directions is 
constrained mainly by the jet data according to Fig.\ \ref{fig:figTwo} or  
Table \ref{table:table5}.

%Conflicts in the global PDF fit, in the cases we have examined here,  were 
% found to be small, compared to the 
% tolerance parameter delta chisqr = 100 for 90 percent confidence, which would 
% correspond to expanding errors by factor of 6.

\begin{acknowledgments}
I am grateful to my late colleague and friend Wu-Ki Tung for the pleasure of 
many discussions on these issues.  
This research was supported by National Science Foundation grant PHY-0354838. 
\end{acknowledgments}

\section*{Appendix: Rediagonalizing the Hessian matrix}

This Appendix describes details of the procedure that simultaneously 
diagonalizes the coordinate dependence of $\chi^2$ and one additional 
quantity within the quadratic approximation.
The procedure was first described in Appendix B of \cite{Collins}, 
but its significance was not recognized in that paper.

The Hessian method is based on the quadratic expansion of $\chi^{2}$ 
in the neighborhood of the minimum that defines the best fit to 
the data:
\begin{equation}
\chi^{\, 2} \, = \, \chi_{0}^{\, 2} \, + 
\sum_{i=1}^N \sum_{j=1}^N H_{i j} \, x_i \, x_j \; ,
\label{eq:app1}
\end{equation}
where $x_i$ is the displacement $a_i - a_i^{(0)}$ from the minimum in the 
original parameter space, and the Hessian matrix is defined by
\begin{equation}
H_{i j} \, = \, \frac{1}{2} 
 \left(\frac{\partial^{\, 2}\chi}{\partial x_i\,\partial x_j}\right)_0 \, .
\label{eq:app2}
\end{equation}
(The Hessian matrix is usually defined without the overall factor 1/2, 
but the normalization used here is more convenient for present purposes.)
Eq.~(\ref{eq:app1}) follows from Taylor series in the neighborhood of the 
minimum.  It contains no first-order terms because the expansion is about the 
minimum, and terms smaller than second order have been dropped according to 
the quadratic approximation.

Since $\mathbf{H}$ is a symmetric matrix, it has a complete set of $N$ 
orthonormal eigenvectors $V_i^{(1)},\dots,V_i^{(N)}$:
\begin{eqnarray}
 \sum_{j=1}^N H_{ij} \, V_j^{(k)} &=& \epsilon_k \, V_i^{(k)} 
                                                   \label{eq:app3a} \\
 \sum_{k=1}^N V_{k}^{(i)} \, V_k^{(j)} &=& \delta_{ij} 
                                                   \label{eq:app3b} \\
 \sum_{k=1}^N V_{i}^{(k)} \, V_{j}^{(k)} &=& \delta_{ij} \; . 
                                                   \label{eq:app3c}
\end{eqnarray}
The eigenvalues $\epsilon_k$ are positive because the best fit 
must be at a minimum of $\chi^2$. 
Multiplying (\ref{eq:app3a}) by $V_m^{(k)}$ and summing over $k$ yields
\begin{equation}
  H_{ij} = \sum_{k=1}^N \epsilon_k \, V_i^{(k)} \, V_j^{(k)} 
  \; .
\label{eq:app4}
\end{equation}
We can define a new set of coordinates $\{y_i\}$ that describe 
displacements along the eigenvector directions:  
\begin{eqnarray}
  S_j &=& 1/\sqrt{\epsilon_j} \label{eq:app5a} \\ 
  W_{ij} &=& V_i^{(j)} \, S_j  \label{eq:app5b} \\
  x_i &=& \sum_{j=1}^N W_{ij} \, y_j \; . \label{eq:app5c} 
\end{eqnarray}
Then
\begin{equation}
  \chi^{\, 2} \, = \, \chi_0^{\, 2} \, + \, \sum_{i=1}^N y_i^{\, 2} \; . 
\label{eq:app6} 
\end{equation}

Any additional function $G$ of the original coordinates $\{a_i\}$ can also be 
expressed in terms of the new coordinates $\{y_i\}$ and
expanded by Taylor series through second order:
\begin{equation}
G \, = \, G_0 \, +  \, 
\sum_{i=1}^N P_i \, y_i \, + \,
\sum_{i=1}^N \sum_{j=1}^N Q_{i j} \, y_i \, y_j \; .
\label{eq:app7}
\end{equation}
The symmetric matrix $\mathbf{Q}$, like $\mathbf{H}$, has a complete set 
of orthonormal eigenvectors $U_i^{(1)},\dots,U_i^{(N)}$:

\begin{eqnarray}
  \sum_{j=1}^N Q_{ij} \, U_j^{(k)} &=& \gamma_k \, U_i^{(k)} \label{eq:app8a} \\
  \sum_{k=1}^N U_{k}^{(i)} \, U_k^{(j)} &=& \delta_{ij} \label{eq:app8b} \\
  \sum_{k=1}^N U_{i}^{(k)} \, U_{j}^{(k)} &=& \delta_{ij} \; , \label{eq:app8c}
\end{eqnarray}
from which it follows that
\begin{equation}
  Q_{ij} = \sum_{k=1}^N \gamma_k \, U_i^{(k)} \, U_j^{(k)} \; .
\end{equation}
Defining new coordinates $\{z_i\}$ by 
\begin{equation}
  z_i \, = \, \sum_{j=1}^N  U_j^{(i)} \, y_j
\end{equation}
now leads to 
\begin{eqnarray}
  \chi^2 &=& \chi_0^{\,2} \, + \, \sum_{i=1}^N z_i^{\, 2} \nonumber \\ 
  G      &=& G_0 \, + \, \sum_{i=1}^N 2 \, \beta_i \, z_i \, + \, 
\sum_{i=1}^N \gamma_i \, z_i^{\, 2} \; , 
\label{eq:MainResult}
\end{eqnarray}
where
\begin{equation}
  \beta_i \, = \, \frac{1}{2} \, \sum_{j=1}^N  U_j^{(i)} \, P_j \; .
\end{equation}
\emph{Hence both $\chi^2$ and $G$ are diagonal in the new coordinates $\{z_i\}$}
in the quadratic approximation.  
Eq.~(\ref{eq:newdiag}), which is the basis of this paper, follows immediately 
from (\ref{eq:MainResult}) by choosing $G$ to be the contribution to $\chi^2$ 
from the subset $\mathbf{S}$ of the data.

Because non-quadratic behavior appears at widely different scales in 
different directions of the original parameter space, and because the 
second-derivative matrices are calculated numerically by finite differences, 
it is actually necessary to compute the linear transformation from the old 
coordinates $\{a_i - a_i^{(0)}\}$ to the new coordinates $\{z_i\}$ by a series 
of iterations \cite{multivariate}.  This is done as follows.  The procedure 
described above yields a coordinate transformation $\mathbf{W}$ defined by 
\begin{equation}
  a_i - a_i^{(0)}  = \sum_{j=1}^N W_{i j} \, z_j \; .
\label{eq:transform}
\end{equation}
The coordinates $\{z_i\}$ can be treated as ``old'' coordinates and the 
above steps repeated to obtain a refined set of elements for the matrix 
$\mathbf{W}$.  This process is iterated a few times to obtain the final 
form of the transformation.  The iterative method is simple to 
program: each iteration begins with an estimate of the desired transformation 
matrix $\mathbf{W}$ in (\ref{eq:transform}) and ends with an improved version 
of $\mathbf{W}$.  One can start with the unit matrix $W_{ij} = \delta_{ij}$ 
and iterate until the matrix $\mathbf{W}$ stops changing.  This procedure has 
been found to converge in all of the applications for which it has been tried.

The distance moved away from the minimum in the original coordinate space is 
given by 
\begin{equation}
 D \, = \, \sum_{i=1}^N (a_i - a_i^{(0)})^2 \, = \,  
 \sum_{i=1}^N  \sum_{j=1}^N \left(\sum_{k=1}^N W_{ki} \, 
                W_{kj}\right) z_i \, z_j \, ,
\end{equation}
which corresponds to the choice 
\begin{equation}
Q_{ij} = \sum_{k=1}^N W_{ki} \, W_{kj}
\end{equation}
in the iterative scheme.  This choice produces eigenvector directions that are 
characterized by how rapidly $\chi^2$ changes in the original parameter space, 
leading to a clear distinction between ``steep directions'' in which $\chi^2$ 
increases rapidly with displacement in the original parameters, and 
``flat directions'' in which the $\chi^2$ increases only slowly.  The degree 
of steepness or flatness is measured by the eigenvalues of $\mathbf{Q}$.  

In the PDF analysis, a large number of free parameters are used in order to 
reduce the ``parametrization error'' caused by the need to represent unknown 
continuous parton distribution functions by approximations having a finite 
number of parameters.  In that application, the logarithms of the eigenvalues 
of $\mathbf{Q}$ are found to be roughly uniformly distributed, with the 
smallest and largest eigenvalues having a huge ratio.  As a result, the 
iterative method has been found to be necessary even to carry out the 
conventional Hessian analysis, where only $\chi^2$ needs to be diagonalized.


\begin{thebibliography}{99}

\bibitem{HessianMethod}
  J.~Pumplin {\it et al.},
  %``Uncertainties of predictions from parton distribution functions. 2. The
  %Hessian method,''
  Phys.\ Rev.\  D {\bf 65}, 014013 (2001)
  [arXiv:hep-ph/0101032];
  %%CITATION = PHRVA,D65,014013;%%

\bibitem{LagrangeMethod}
  D.~Stump {\it et al.},
  %``Uncertainties of predictions from parton distribution functions. 1. The
  %Lagrange multiplier method,''
  Phys.\ Rev.\  D {\bf 65}, 014012 (2001)
  [arXiv:hep-ph/0101051].
  %%CITATION = PHRVA,D65,014012;%%

\bibitem{cteq66}
  P.~M.~Nadolsky {\it et al.},
  %``Implications of CTEQ global analysis for collider observables,''
  Phys.\ Rev.\  D {\bf 78}, 013004 (2008)
  [arXiv:0802.0007 [hep-ph]].
  %%CITATION = PHRVA,D78,013004;%%

\bibitem{CT09}
 J.~Pumplin, J.~Huston, H.L.~Lai, Wu-Ki Tung and C.-P.~Yuan,
``Collider Inclusive Jet Data and the Gluon Distribution,''
  [arXiv:0904.2424 [hep-ph]].

\bibitem{MSTW08}
  A.~D.~Martin, W.~J.~Stirling, R.~S.~Thorne and G.~Watt,
  %``Parton distributions for the LHC,''
  arXiv:0901.0002 [hep-ph].
  %%CITATION = ARXIV:0901.0002;%%

\bibitem{alternatives}
The sum of quadratic deviations in Eq.\ (\ref{eq:chisqdef}) is the natural 
measure of fit quality if the experimental errors are Gaussian-distributed.  
If they are not Gaussian, alternative forms might be worth considering in 
which extreme values of the deviation $\Delta_i \equiv (D_i - T_i)/E_i$ are 
assigned more weight (e.g., $\chi^2 = \sum_i \Delta_i^{\, 4}$) to force the 
fit toward describing every data point satisfactorily; or more likely less 
weight (e.g., $\chi^2 = C \, \sum_i \log(1 + \Delta_i^{\, 2}/C)$) to downplay 
the influence of extreme outlying points.

\bibitem{Collins}
  J.~C.~Collins and J.~Pumplin,
  ``Tests of goodness of fit to multiple data sets,''
  arXiv:hep-ph/0105207.
  %%CITATION = HEP-PH/0105207;%%

\bibitem{multivariate}
  J.~Pumplin, D.~R.~Stump and W.~K.~Tung,
  %``Multivariate fitting and the error matrix in global analysis of data,''
  Phys.\ Rev.\  D {\bf 65}, 014011 (2001)
  [arXiv:hep-ph/0008191].
  %%CITATION = PHRVA,D65,014011;%%

\bibitem{E605}
  G.~Moreno {\it et al.},
  %``Dimuon production in proton - copper collisions at $\sqrt{s}$ = 38.8-GeV,''
  Phys.\ Rev.\  D {\bf 43}, 2815 (1991).
  %%CITATION = PHRVA,D43,2815;%%

\bibitem{CDFR2}
  T.~Aaltonen {\it et al.}  [CDF Collaboration],
  %``Measurement of the Inclusive Jet Cross Section at the Fermilab Tevatron
  %p-pbar Collider Using a Cone-Based Jet Algorithm,''
  Phys.\ Rev.\  D {\bf 78}, 052006 (2008)
  [arXiv:0807.2204 [hep-ex]].
  %%CITATION = PHRVA,D78,052006;%%

\bibitem{D0R2}
  V.~M.~Abazov {\it et al.}  [D0 Collaboration],
  %``Measurement of the inclusive jet cross-section in $p \bar{p}$ collisions at
  %$s^{91/2)}$ =1.96-TeV,''
  Phys.\ Rev.\ Lett.\  {\bf 101}, 062001 (2008)
  [arXiv:0802.2400 [hep-ex]].
  %%CITATION = PRLTA,101,062001;%%

\end{thebibliography}
\end{document}